\documentclass[11pt, nobibnotes, prb,amsmath]{revtex4}
\usepackage{amsmath, amsthm, amssymb}
\newtheorem{theorem}{Theorem}
\newtheorem{lemma}{Lemma}

\newcommand{\beqn}{\begin{equation}}
\newcommand{\eeqn}{\end{equation}}

\newcommand{\veps}{\varepsilon}

\begin{document}

\title{Stability of Three Unit Charges. Necessary Conditions.}

\author{D.K. Gridnev}
\email[Electronic address: ]{dima_gridnev@yahoo.com}
\author{C. Greiner}
\author{W. Greiner}
\affiliation{Institut f{\"u}r Theoretische Physik,
Robert-Mayer-Str. 8-10, D--60325 Frankfurt am Main, Germany}
\begin{abstract}
We consider the stability of three Coulomb charges $\{ +1, -1 , -1
\}$ with finite masses in the framework of nonrelativistic quantum
mechanics. A simple physical condition on masses is derived to
guarantee the absence of bound states below the dissociation
thresholds. In particular this proves that certain negative muonic
ions are unstable, thus extending the old result of Thirring
\cite{thirring} to the actual values of all masses. The proof is
done by reducing the initial problem to the question of binding of
one particle in some effective potential.
\end{abstract}

\pacs{Give the pacs here}

\maketitle

\section{Introduction}\label{sec-intro}
The stability of three particles with pure Coulomb forces is an
old and extensively studied problem, this is to explain why
certain ions and molecules stay as a whole and some dissociate
into a bound pair and a single particle. Under stability of the
Hamiltonian $H$ we shall understand the existence of a bound state
with the energy strictly less than $\inf \sigma_{ess} (H)$, {\it
i.e.} a stationary state below all dissociation thresholds. As a
quantum system three particles with Coulomb interactions
demonstrate interesting behavior. It is known that three charges
$\{ 1 + \veps , -1, -1 \}$ for any $\veps
>0$ form the system, which is stable regardless of mass values.
However at $\veps = 0$ the situation abruptly changes and due to
the screening effect not all systems remain stable. The typical
example is the unstable muonic hydrogen ion $p \mu^- e^- $, where
the heavy muon is tightly bound to the proton and screens the
positive charge for the electron. This interesting effect is well
studied, in particular it has been proved \cite{hill} that for
equal dissociation thresholds the system remains stable
(Refs.~\onlinecite{martin1,martin3} explain very well how
stability depends on masses and charges).

Among the instability proofs for three unit charges the most
appealing is that of Thirring\cite{thirring}, which does not
require any numerical calculations. Thirring considered a negative
hydrogen ion with an infinitely heavy proton, and proved  that
such negative ion is unstable when its second negatively charged
particle is lighter than electron by a factor larger than $\pi$
($\pi$ can be replaced \cite{martin3,glaser} with a better
constant $1.57$). In his method Thirring exploited the fact that
the ground state in the hydrogen atom is substantially separated
from other states in its spectrum (the non-degenerate energy
levels have $1/ n^2$ dependence), and thus its role becomes
emphasized. This suggests an idea to move the problem to the space
spanned by the projector $P_0 = | \phi_0 \rangle \langle \phi_0 |
\otimes 1$, where $\phi_0$ is the ground state of Hydrogen. After
estimating the part of the repulsion, which is present in this
space, the problem reduces to checking the binding of one particle
in some effective potential. The contribution from the attractive
interaction term coming from additional particle is easy to treat
because it commutes with $P_0$. Yet this is no longer true if one
considers particles of finite mass or any system of four
particles.

Thirring's bound was improved \cite{martin1} and it was shown that
the muonic ion $p \mu^- e^- $ is unstable for actual values of all
masses. However this extension (Eq.~24 in
Ref.~\onlinecite{martin1}) is weak in the sense that it fails when
the second particle is heavy compared to other particles. (For a
physical example, it does not prove that the ion $\mu^- p e^+ $ is
unstable, yet we shall prove it here). This extension still uses
Thirring's treatment of repulsion and it is unclear how one could
extend it to four particles. Armour\cite{armour} with his method
proved the instability of such systems as positron-hydrogen-atom
$e^- p e^+$ and $\mu^-  p e^+ $, but this method requires certain
numerical assistance. It also bases on the separation of variables
in the problem of two fixed centers, which makes it inapplicable
to four particles because the variables do not separate even in
the case of three fixed centers.

Here we follow the Thirring's idea but the nucleus does not have
to be infinitely heavy. The derived physical condition restricts
the ratio of Jacobi masses, which makes the system stable. It can
be used in conjunction with Thirring's result and convexity
properties of stability curve \cite{martin1} for a reasonable
determination of stability area free from any numerics. The
present method has an important advantage in that it admits
generalization to four particles. It should be mentioned that both
Thirring's method and the new one share the same deficiency,
namely, they are not applicable when the dissociation thresholds
in the system are close or equal ({\it e.g.} when two like charges
have equal masses). In particular, using such methods one cannot
prove the ``overheating'' effect, when a system with charges $\{
+1, -q,-q \}$ and equal masses loses its binding for some $q>1$.
Nowadays various charged particles are produced in laboratories
and it is, of course, of interest to know the principles behind
formation of exotic atoms or molecules. All this motivates the
present analysis.

\section{From Three Particles to One in Effective Potential}\label{sec-2}
Let $m_i , q_i, {\bf r}_i \in \mathbb{R}^3 $ denote masses,
charges and position vectors of particles $i = 1,2,3$. We put $q_1
= +1$, and $q_{2,3} = -1$, and the interactions between the
particles are $V_{ik} = q_i q_k / |{\bf r}_i - {\bf r}_k |$. We
enumerate the particles in such a way, that the particles $(1,2)$
form the lowest dissociation threshold. The stability problem with
Coulomb interactions is invariant with respect to scaling all
masses \cite{martin1}, so we can put $\hbar =1$. We separate the
center of mass motion in the Jacobi frame \cite{messiah} putting
${\bf x} = {\bf r}_1 - {\bf r}_2$ and ${\bf y } = {\bf r}_3 - {\bf
r}_1 + a {\bf x}$, where $a = m_2 /(m_1 + m_2 )$ is the mass
parameter invariant with respect to mass scaling. The reduced
masses and Jacobi momenta are respectively $\mu_x = m_1 m_2 / (m_1
+ m_2 )$, $\mu_y = m_3 (m_1 + m_2 ) / (m_1 + m_2 + m_3 )$ and
$p_{x,y} = -i\nabla_{x,y}$. It is convenient to scale all masses
so that $\mu_x = 2$. (In Sec.~\ref{summary} we shall rescale them
back). The Hamiltonian for the system on the tensor product space
$L_2 (\mathbb{R}^3) \otimes L_2 (\mathbb{R}^3) $ is
\begin{equation}\label{ham0}
      H = h_{12} \otimes 1 + 1\otimes \frac{p_{y}^2}{2 \mu_y} +
      W \nonumber
\end{equation}
where
\begin{equation}\label{W0}
W  = V_{13} + V_{23} = - \frac 1{|a {\bf x} - {\bf y}|} +
\frac{1}{| (1-a) {\bf x} + {\bf y}|}
\end{equation}
and $h_{12} = p_{x}^2 /4 - 1/x$ is the Hamiltonian of the pair of
particles (1,2) (notation $x$ is used instead of $|{\bf x}|$). The
ground state wave function of $h_{12}$ is $\phi_0 = \sqrt{8/ \pi}
\exp(-2x)$ so that $h_{12} \phi_0 = -\phi_0$. The Hamiltonian $H$
is self-adjoint on $\mathcal{D} (H) = H^2 (\mathbb{R}^6)$ (square
integrable functions having partial derivatives up to the second
order in the weak distributional sense) and by the HVZ theorem
$\sigma_{ess} ( H ) = [ -1 , \infty) $. We split positive and
negative parts of $W$ by introducing $W_- : = (|W| - W)/2$ and
$W_+ := (- W)_- = (|W| + W)/2$ and we have $W = W_+ - W_-$, where
$W_{\pm} \geq 0$. Instead of $H$ we shall consider the Hamiltonian
\begin{equation}\label{ham}
    \tilde H = h_{12} \otimes 1 + 1\otimes \frac{p_{y}^2}{2 \mu_y}
    - W_-
\end{equation}
Note that the part $W_-$ can also be expressed as
\begin{equation}\label{W_-}
W_- =  - ( V_{13} +  V_{23} F({\bf x}, {\bf y}) )
\end{equation}
where $F = 1$ when $|V_{13}| \geq |V_{23}|$ and $F = |V_{13}| /
|V_{23}|$ when $|V_{13}| \leq |V_{23}|$, and we have $0 \leq F
\leq 1$. Because $\| V_{23} F \phi \| \leq \| V_{23} \phi \|$ we
can directly apply Kato's theorem \cite{simon} on self-adjointness
of atomic Hamiltonians and find out that $\tilde H$ is
self-adjoint on $\mathcal{D} ( \tilde H ) = \mathcal{D} ( H )$. We
cannot though directly apply the HVZ theorem to locate $\inf
\sigma_{ess} ( \tilde H )$ but we observe that the following
inequality holds $(H - V_{23}) \leq \tilde H \leq H$. Here
$(H-V_{23} )$ is the original Hamiltonian without repulsion, which
is bounded from below and by the HVZ theorem $\inf \sigma_{ess} (
H - V_{23} ) = -1$. Thus from the min-max principle \cite{simon}
we get $\inf \sigma_{ess} ( \tilde H ) = -1$.

Now let us assume that $H $ is stable, {\it i.e.} $H$ has a bound
state with the energy less than $-1$.  Because $\tilde H \leq H$
from the variational principle we conclude that $\tilde H$ also
has a bound state $\Psi \in \mathcal{D} (H )$ with the energy
below $\inf \sigma_{ess} ( \tilde H ) = -1$ which means
\begin{equation}\label{binding}
    \tilde H \Psi = (-1-\delta) \Psi
\end{equation}
where $\delta > 0$ is the extra binding energy. Let us introduce a
projection operator $P_0 = | \phi_0 \rangle \langle \phi_0 |
\otimes 1$ $( P_0 : \mathcal{D} ( H ) \rightarrow \mathcal{D} ( H
))$ and put $\eta := P_0 \Psi$ and $\xi := (1 - P_0 ) \Psi$, where
obviously $\eta \bot \xi$ and $\Psi = \eta + \xi$ and $\eta, \xi
\in \mathcal{D} ( H )$. Taking the scalar product of each side of
(\ref{binding}) with $\eta$ and $\xi$ we obtain
\begin{eqnarray}
  \langle \eta | 1\otimes \frac{p_{y}^2}{2\mu_y} - W_- | \eta\rangle -  \langle
  \eta | W_- |  \xi \rangle &=& -\delta \| \eta \| ^2 \label{eq1}\\
 \langle \xi | h_{12} \otimes 1 |  \xi \rangle  +  \langle \xi | 1\otimes
 \frac{p_{y}^2}{2\mu_y} - W_- | \xi \rangle -  \langle
  \xi | W_- |  \eta \rangle &=& (-1-\delta) \| \xi \|^2 \label{eq2}
\end{eqnarray}
where we have used $\langle \eta | 1 \otimes p_{y}^2 | \xi \rangle
= 0$ because $P_0$ and $1 \otimes p_{y}^2 $ commute. We shall
assume that $\| \eta \| , \| \xi \| \neq 0$ (we shall get rid of
this assumption in Theorem~\ref{myth}), then we are free to choose
such normalization of $\Psi$ that $\| \xi \| = 1$.

From the bound spectrum of $h_{12}$ we have \cite{thirring}
$h_{12}\otimes 1 \geq -P_0 - 1/4 (1-P_0 )$, hence for the first
term in (\ref{eq2}) we get the bound $\langle \xi | h_{12} \otimes
1 | \xi \rangle \geq - 1 /4$. Introducing two non-negative
constants $\alpha :=\sqrt{\langle \eta | W_- | \eta \rangle }$ and
$\beta := \sqrt{\langle \xi | W_- | \xi \rangle}$ we get by virtue
of the Schwarz inequality $|\langle \xi | W_- | \eta \rangle |
\leq \alpha \beta$. Now we can rewrite
Eq.~(\ref{eq1})--(\ref{eq2}) to obtain the main pair of
inequalities
\begin{eqnarray}
  \langle \eta | 1\otimes \frac{p_{y}^2}{2\mu_y} - W_- | \eta\rangle -  \alpha
  \beta &<& 0 \label{1}\\
 \langle \xi | 1\otimes
 \frac{p_{y}^2}{2\mu_y} - W_- | \xi\rangle -  \alpha
  \beta &<& - \frac34  \label{2}
\end{eqnarray}
Using the second inequality we shall find $\max \beta /\alpha $
and by substituting this value into (\ref{1}) we shall formulate
the stability condition.

\begin{lemma}\label{lemm} Suppose that
Eq.~(\ref{2}) holds and $\mu_y < 3/2$, then the following
inequality is true
\begin{equation}\label{ratio}
    \beta < \left( \sqrt{\frac3{2\mu_y }} - 1 \right)^{-1} \alpha
\end{equation}
\end{lemma}
\begin{proof}
First, let us show that for $A \geq 0$
\begin{equation}\label{l1}
     \inf_{\substack{\chi \in \mathcal{D} (H) \\ \|\chi\|=1}}
     \langle \chi | 1\otimes \frac{p_{y}^2}{2\mu_y} - A W_- | \chi
     \rangle \geq - \frac{A^2 \mu_y }{2}
\end{equation}
It suffices to prove this for $\chi \in C_{0}^\infty (\mathbb{R}^6
)$. Using (\ref{W_-}) from the variational principle we get
\begin{gather}
    \langle \chi | 1\otimes \frac{p_{y}^2}{2\mu_y} - A W_- | \chi
     \rangle \geq \int d {\bf x} \int d {\bf y} \chi^*
    ({\bf x} , {\bf y}) \left( \frac{p_{y}^2}{2\mu_y} -
     \frac A{|a{\bf x} - y|} \right) \chi ({\bf x} , {\bf y}) \nonumber \\\geq
     - \frac{A^2 \mu_y }2 \int d{\bf x} \int d {\bf y} |\chi|^2
     ({\bf x} , a{\bf x} +{\bf y})  = -  \frac{A^2 \mu_y }{2} \| \chi \|^2 \label{integrals}
\end{gather}
from which (\ref{l1}) follows and where we have used the explicit
expression for the ground state energy of the hydrogen atom.
(Using an appropriate set of trial functions it is easy to show
that there is an equality sign in (\ref{l1}), but we do not need
this for our purposes). Now using (\ref{l1}) we obtain the
following chain of inequalities
\begin{gather}
       \inf_{\substack{\xi \in \mathcal{D} (H) \\ \| \xi\| =1 \\
       \langle \xi | W_- | \xi
       \rangle = \beta^2 }} \langle \xi | 1 \otimes
       \frac{p_{y}^2}{2\mu_y} - W_- | \xi \rangle =  \max_{\lambda \geq -1
       } \inf_{\substack{\xi \in \mathcal{D} (H) \\ \| \xi\| =1 \\ \langle \xi | W_- | \xi
       \rangle = \beta^2 }} ( \langle \xi | 1 \otimes
       \frac{p_{y}^2}{2\mu_y} - (\lambda + 1 )W_- | \xi \rangle +
       \lambda \beta^2 ) \nonumber \\ \geq  \max_{\lambda \geq -1
       } \inf_{\substack{\xi \in \mathcal{D} (H) \\ \| \xi\| =1}} ( \langle \xi | 1 \otimes
       \frac{p_{y}^2}{2\mu_y} - (\lambda + 1 )W_- | \xi \rangle +
       \lambda \beta^2 ) \geq \max_{\lambda \geq -1
       }  ( \lambda \beta^2 - (\lambda + 1 )^2 \mu_y
       /2 ) \nonumber \\ = \frac{\beta^4 }{2 \mu_y } - \beta^2
       \nonumber
\end{gather}
Substituting this result into (\ref{2}) and putting $\alpha = s
\beta$ we obtain
\begin{equation}\label{ratio2}
         \frac{\beta^4 }{2 \mu_y } - (s +1 ) \beta^2 < - \frac34
\end{equation}
Now simply minimizing the left-hand side of (\ref{ratio2}) over
$\beta ^2 $ we obtain the lower bound on $s$, which gives us
Eq.~(\ref{ratio}). \end{proof}

It is worth noting that the relation in Lemma is a version of the
uncertainty principle, when $\beta^2 $ grows, the kinetic energy
term grows faster like $\beta^4$.  Let us introduce an effective
potential $V_{eff} (y) := \int d {\bf x} |\phi_0 |^2 W_- $. We
formulate the result as
\begin{theorem}\label{myth}
If the system of three charges is stable and $\mu_y < 3/2$ then
the particle with mass $\mu_y$ must have a bound state in the
potential $-\left( 1 + (\sqrt{3 /2\mu_y } - 1)^{-1}\right)
V_{eff}$.
\end{theorem}
\begin{proof}
We have $\| \eta \| , \| \xi \| \neq 0$. The function $\eta $ has
the factorized form $\eta = \phi_0 ({\bf x}) f ({\bf y})$, where
$f \in H^2 (\mathbb{R}^3 )$, $\| f \| \neq 0$. By substituting
(\ref{ratio}) into (\ref{1}) and using expressions for $\alpha^2 $
and $\eta$ we get the necessary condition for stability
\begin{equation}\label{stab1}
\langle f | \frac{p_{y}^2}{2 \mu_y} - \left(1 + ( \sqrt{3/2 \mu_y
} - 1 )^{-1} \right) V_{eff} | f \rangle < 0
\end{equation}
In the next section we shall study $V_{eff} (y)$ and show that it
is a continuous function decaying like $1/ y^2$. Inequality
(\ref{stab1}) means that a particle having mass $\mu_y$ has a
bound state in this potential.

Now let us complete the proof considering the case when either $
\xi  =0$ or $ \eta = 0$. If $ \xi = 0$ we have $\beta = 0$ and
substituting this into (\ref{1}) we get a condition more stringent
than Eq.~(\ref{stab1}). If $\eta = 0$ we have $\| \xi \| = 1 $ and
$\alpha = 0$, substituting this into (\ref{2}) and using
(\ref{l1}) for $\mu_y < 3/2$ results in the contradiction.
\end{proof}

\section{Binding in Effective Potential}
In this section we shall analyze the effective potential and find
out at which values of the coupling constant $\lambda$ the
Hamiltonian $p_{y}^2 - \lambda V_{eff}$ may have bound states. It
turns out that the effective potential in our case has a
nonphysical term, which is a long-range attraction of the type $1/
y^2$. This nonphysical behavior stems from cutting off the
positive part of the potential and results in the infinite number
of bound states at the point of binding (that is why it is meant
nonphysical). But as it is well-known since the result of Hilbert
and Courant \cite{courant}, even this long-range attraction does
not guarantee binding, for $ \lambda \max_y y^2  V_{eff} (y) \leq
1/4$ the inequality $p_{y}^2 - \lambda V_{eff} \geq 0$ holds, {\it
i.e.} no binding occurs, (see also the proof in
Ref.~\onlinecite{simon}, vol. 2). Thus the non-trivial critical
coupling constant exists, and we have to determine it. On the
other hand, in such potentials short-range repulsive terms do not
play any role for binding \cite{1/r2}. It would be of interest to
get rid of this nonphysical behavior in the future (pay attention
that the Thirring's effective potential \cite{thirring} behaves at
infinity like $1/y^3$!).

To calculate $V_{eff}$ we must cut off the positive part of $W$.
From (\ref{W0}) $W\leq 0$ is equivalent to $\cos \theta \geq
x/(\omega y)$, where $\omega = (a-1/2)^{-1}$ and $\cos \theta =
{\bf x} \cdot {\bf y} / xy$. We shall consider separately two
cases $a > 1/2$ and $a \leq 1/2$.

\subsection{Case when $a > 1/2$}
The integration is simpler in this  case. After direct integration
in spherical coordinates over the area where $\cos \theta \geq x/(
\omega y)$ we obtain
\begin{gather}
    V_{eff} (y) = 16 y ^2 \int_{0}^\omega d s \: s e^{ -4sy}  \left(
    \frac{\sqrt{a(1-a)s^2 + 1}}{a(1-a)} - \frac{|as-1|}{a}
    - \frac{(1-a)s + 1}{1-a}  \right) \nonumber \\
    = \frac{16 y ^2}{a(1-a)} \int_{0}^\omega d s \: s e^{ -4sy} (\sqrt{a(1-a)s^2 + 1}
    -1) + U \label{veff1}
\end{gather}
where
\begin{equation}\label{u}
     U = 32 y^2 \int_{1/a}^\omega d s \: s e^{ -4sy}  ( 1/a - s) <0
\end{equation}
Now we do not have to carry out integration in (\ref{u}), it is
enough to see that it is a short-range repulsion, which does not
play a role in our case. To calculate the first term in
(\ref{veff1}) we use $\sqrt{a(1-a)s^2 + 1} \leq 1 + a(1-a)s^2 /2 $
to get
\begin{equation}\label{veff2}
    V_{eff } < 8 y^2 \int_{0}^\omega d s \: s^3 e^{ -4sy}  <
    \frac 3{16 y^2 }
\end{equation}
where after the integration we have dropped the short-range
negative terms. Finally, we have $p_{y}^2 - \lambda V_{eff} \geq
p_{y}^2 - \lambda (3/16) y^{-2}$. The following inequality
\cite{simon} holds $p_{y}^2 -  (1/4) y^{-2} \geq 0$. Thus in the
case of binding, {\it i.e.} when such $f$ exists that $\langle f |
p_{y}^2 - \lambda V_{eff} | f \rangle < 0$, we must have $\lambda
>
 4/3$. Comparing this with (\ref{stab1}) we obtain that three
 charges form unstable system if $\mu_y < 3/2$ and
 \begin{equation}\label{same}
    2 \mu_y \left(1 + ( \sqrt{3/2 \mu_y
} - 1 )^{-1} \right) < 4/3
\end{equation}
Solving this simple inequality tells us that the system is
unstable when $\mu_y < 2(11- 2 \sqrt{10})/27 \simeq 0.3463$.

\subsection{Case when $a \leq 1/2$}
First let us take $a< 1/2$. We shall write $W(a)$ instead of $W$
to point out the dependence on parameter $a$. We can alleviate the
integration noting that $W(a) = - W(1-a)$, thus we have $W_- (a) =
(-W(1-a))_- = W_+ (1-a) = W(1-a) + W_- (1-a)$. From this we
conclude $V_{eff} (a) = - \overline{W } (a) + V_{eff} (1-a)$. The
additional integral $\overline{W } (a) := \int d{\bf x} |\phi_0
|^2 W$ is easy to calculate and $V_{eff} (1-a)$ for $a < 1/2$ we
have already calculated. We obtain
\begin{gather}
    - \overline{W} (a) = 16 y^2 \int_{0}^\infty ds \: e^{ -4sy}  s
    \left( \left( \frac 1a - \frac 1{1-a} \right) + \left| s - \frac
    1{1-a} \right| - \left| s - \frac
    1a \right| \right) \nonumber \\
    = 32 y^2 \int_{1/(1-a)}^{1/a} ds \: e^{ -4sy}  \left( s^2 - \frac
    s{1-a} \right)
    + 32 y^2 \int_{1/a}^{\infty} ds \: e^{ -4sy}  s \left( \frac 1{a} -
    \frac 1{1-a} \right) \label{ovW}
\end{gather}
Using (\ref{veff1}),~(\ref{u}) and approximation for the square
root $\sqrt{a(1-a)s^2 + 1} \leq 1 + a(1-a)s^2 /2 $ gives us
\begin{equation}\label{veff6}
    V_{eff} (1-a) \leq 8 y^2 \int_{0}^{-\omega} ds \: s^3 e^{ -4sy}  + 32
    y^2 \int_{1/(1-a)}^{-\omega} ds \: e^{ -4sy}  \left( \frac s{1-a} -
    s^2 \right)
\end{equation}
Summing (\ref{ovW}) and (\ref{veff6}) and calculating the
integrals explicitly gives us the following expression for $a <
1/2$
\begin{gather}\label{veff100}
    V_{eff} <  \frac 3{16 y^2 } - e^{-4y/a} \left[ \frac
    {8(1-a)y}{a^2} + 2 \frac{2-a}a + \frac 1y \right] \nonumber \\ - e^{4\omega y}
    \left[ -2y \omega (\omega +2)^2
    + ( 3 \omega  /2  + 1)(\omega +2 ) - \frac 1y (3\omega /4 + 1) + \frac 3{16y^2}\right]
\end{gather}
For $a < 1/2$ we have $\omega < -2 $, and it is easily seen that
all terms in square brackets are positive (this leads again to
short-range potentials), meaning that $V_{eff} < (3/16) y^{-2}$,
which gives the same condition for stability as (\ref{same}). We
do not consider explicitly $a= 1/2$, it is done analogously and
also results in (\ref{same}).

\section{summary}\label{summary}
We have initially scaled all masses $m_i \rightarrow 2 m_i /
\mu_x$, making $\mu_x = 2$. Now rescaling it back we get through
(\ref{same}) that the system of three charges is unstable if
$\mu_y / \mu_x < (11- 2 \sqrt{10})/27 \simeq 0.1732$. In the case
of infinitely heavy nucleus this is $m_3 / m_2 < 0.1732$, which is
worse than the refined \cite{martin1,glaser} Thirring's estimate
$m_3 / m_2 < 1/ 1.57$. The accuracy is lost at the point of
cutting the positive part of the potential, which induces a
long-range attraction. However this is more than enough to prove
that the muonic ions $p \mu^-  e^- $ or $\mu^- p e^+ $ are
unstable for the actual values of all three masses. The case of
four unit charges $\{+ 1 , +1 , -1, -1 \}$ is treated similarly
but the calculations are more involved \cite{gridnev} and results
would be published elsewhere. Let us also stress that the obtained
condition is physical. Both Jacobi masses determine Bohr radii for
the particle orbits, the orbit within the pair of particles (1,2)
and the orbit for the third particle in the field of this pair
with respect to the pair's center of mass. If the orbit of one
negative particle is outdistanced then the attraction from the
positive charge is screened off by  the other negative particle
and the system becomes unbound.

\end{document}